\documentclass[showpacs,preprintnumbers,amsmath,amssymb,floatfix,nofootinbib]{revtex4}

\usepackage{color}   
\usepackage{cancel}
\usepackage{color}
\usepackage{tikz}
\usetikzlibrary{decorations.pathmorphing,decorations.markings,trees}
\usepackage[utf8]{inputenc}
\usepackage[T1]{fontenc}
\usetikzlibrary{shapes,arrows,positioning,automata,backgrounds,calc,er,patterns}
\usepackage{tkz-graph}
\tikzstyle{vertex}=[circle, draw, inner sep=0pt, minimum size=6pt]

\usepackage{graphicx}
\usepackage{dcolumn}
\usepackage{float}
\usepackage{bm}

\usepackage[utf8]{inputenc}
\usepackage[T1]{fontenc}
\usepackage{ulem}

\begin{document}

\def\bb    #1{\hbox{\boldmath${#1}$}}
\def\bb    #1{\hbox{\boldmath${#1}$}}

\def\blambda{{\hbox{\boldmath $\lambda$}}}
\def\eeta{{\hbox{\boldmath $\eta$}}}
\def\bxi{{\hbox{\boldmath $\xi$}}}
\def\bzeta{{\hbox{\boldmath $\zeta$}}}
\def\sD{D \!\!\!\!/}
\def\sd{\partial \!\!\!\!/}
\def\EQ{{\hbox{\boldmath $Eq.(\ref$}}}
\def\qcd{{{}^{\rm QCD}}}   
\def\qed{{{}^{\rm QED}}}   
\def\2d{{{}_{\rm 2D}}}         
\def\4d{{{}_{\rm 4D}}}         

\today

\large

\title{ Anomalous fermions}

\author{A. V. Koshelkin}

\affiliation{National Research Nuclear University MEPhI, 
Kashirskoye shosse, 31, 115409 Moscow, Russia}

\begin{abstract}
The  Dirac-like equation governing dynamics of free anomalous fermions is derived. The basis bispinors controlling the obtained solutions of this equation turn out to be normalized by the area confining  a region in the bispinor Clifford geometric  space, rather than by the Dirac scalar product, as it takes place in the case of the standard fermions. Therewith, the area value  of this region is found to be equal to the double fermion mass. 
 Quantizing the  solutions of such an equation is shown to support  the anticommutative relations for the studied   fermions and antifermions fields with necessity.    
The  discrete symmetries of the obtained fermion  fields  are studied in detail, They  turn out to be opposite  to the C, P, T symmetries of the standard Dirac fermions in many cases, while the Lagrangian governing  the dynamics of the anomalous fermions is found to be invariant as far as the  C, P, T symmetries independently. The dark matter problem is discussed in the context of the anomalous fermions.
\end{abstract}


\maketitle

\section{Introduction}

The Dirac equation discovered on 1928\cite{Dir28} plays a key role in physics since it has been revealed. The solutions of such an equation are studied in detail, and  are entirely described in a lot of textbooks (see, for example\cite{Itz,Pes95}). The classic  Dirac equation is unconditionally and repeatedly verified by studying all processes where $1/2$-spin fermions participate. However, the problem of searching for new particles still   has been actual  in the context of both studying dark matter and the unification of interactions.

In the present paper a new Dirac-like equation governing $1/2$-spin particles is obtained and studied in detail. The derived solutions of this  equation are shown to support   the necessity of quantizing the anomalous  fermion fields  according to the Fermi-Dirac  rule. Therewith, the basis bispinor governing evolution of the derived solution in the bi-spinor space is found to be normalized by means of the area in the Clifford geometric space rather than by the Dirac scalar products. The discrete $C,P,T$ symmetries of the obtained solutions are studied. They turn out to be opposite to the ones taking place for  the classical fermion fields, mainly in the cases of the $P$ and $T$ symmetries.

 We obtained the  Lagrangian corresponding to the  motion equations of the anomalous fermions which   turns out to be invariant with respect to $C,P,T$,  separately. Based on this  Lagrangian the standard and axial current are derived, and are studied with regards to conservation.  The anomalous fermions are discussed in the context of the candidates for dark matter particles.

 The paper consists of introduction, a single main section, conclusion and an appendix.

\section{The Dirac-like equations for anomalous fermions}

We start from the Klein-Gordon equation for the scalar field $\phi (x)$ whose  mass is $m$, 

\begin{eqnarray}\label{eq0}
(\partial_\mu \partial^\mu +m^2)\phi (x) =0.
\end{eqnarray}

Provided that $\phi (x)$ is one of the component of the bispinor field $\psi (x)$,  we obtained, partially following the Dirac idea\cite{Dir28},

\begin{eqnarray}\label{eq1}
((p \gamma)\mp i\gamma^5m)  ((p \gamma)\mp i\gamma^5m)\psi (x) =0, 
\end{eqnarray} 
where $p_\nu = i \partial_\nu$, whereas $\gamma^\nu$ and $\gamma^5$ are the Dirac matrices which we take in the Weyl representation\cite{Pes95}.

\begin{eqnarray}
 \label{eq2}
  \gamma^0   = 
\begin{pmatrix}
0& I \\
I  & 0  \\
\end{pmatrix} ,
 ~~~~ {\bb \gamma }  = 
 \begin{pmatrix}
0  &- \bb \sigma\\
\bb \sigma  &  0  \\
\end{pmatrix}, 
 \gamma^5   = 
 \begin{pmatrix}
-I& 0 \\
0 & I \\
\end{pmatrix} .
\end{eqnarray}

When the upper sign  in Eq.(\ref{eq1})  governs the field $\psi (x) $, the down sign equitation controls the evolution of the Dirac conjunctive field $\bar \psi (x) $

\begin{subequations}\label{eq3}
\begin{eqnarray}\label{eq3a}
&&((p \gamma) - i\gamma^5m) \psi (x) =0, \\
&&\bar\psi (x) ((p \gamma) + i\gamma^5m) =0,\label{eq3b}
\end{eqnarray} 
\end{subequations}
where $\bar \psi (x)= \psi^\dag (x) \gamma^0$ .

\subsection{The General solution pf the Dirac-like equations}

To derive the solutions of the Eq.({\ref{eq3a}), which we take to be leading in the pair of Eqs.({\ref{eq3a}),({\ref{eq3b}), we go to the momentum representation, expanding $\psi (x)$ and $\bar \psi (x)$ over whole set of plane waves.  As a result, we get

\begin{eqnarray}
 \label{a1}
 \begin{pmatrix}
im  &\omega+\bb p \bb \sigma\\
\omega - \bb p \bb \sigma  & -im   \\
\end{pmatrix} 
 \begin{pmatrix}
\varphi \\
\chi   \\
\end{pmatrix} =0,~~~~\psi (p)= \begin{pmatrix}
\varphi  (p)\\
\chi (p)  \\
\end{pmatrix},
\end{eqnarray}
where $(\omega, \bb p)$ determines a point in the momentum space, $\bb \sigma$ are the Pauli matrices, $\varphi$ and $\chi$ are spinors which are the components of a bispinor $\psi (x)$ in the momentum representation. Following the standard way\cite{Itz,Pes95} we derive the general solutions of Eqs.(\ref{eq3a}),(\ref{eq3b}),  which are  given  by  formulas (see Appendix A)

\begin{eqnarray}
 \label{eq4}
&& \psi(x) =\sum_s \int \frac{d^3\bb p}{\sqrt{2 \varepsilon (\bb p)}(2\pi)^3}  \left( a_s (\bb p)  u_s (\varepsilon (\bb p), \bb p)\exp{(-ipx)} +b^\dag_s (\bb p)  v_s (\varepsilon (\bb p), -\bb p)\exp{(+ipx)}\right) 
\end{eqnarray}

\begin{eqnarray}
 \label{eq5}
&&\bar \psi(x) =\sum_s \int \frac{d^3\bb p}{\sqrt{2 \varepsilon (\bb p)}(2\pi)^3}  \left( a^\dag_s (\bb p)  \bar u_s (\varepsilon (\bb p), \bb p)\exp{(+ipx)} +b_s (\bb p) \bar  v_s (\varepsilon (\bb p), -\bb p)\exp{(-ipx)}\right) ,
\end{eqnarray}
where $p= (\varepsilon (\bb p), \bb p)$ is the  4-momentum of a particle (antiparticle); $\varepsilon (\bb p)= +\sqrt{\bb p^2 + m^2}$ is the fermion (antifermion) energy; $a_s (\bb p)$ and $a^\dag_s (\bb p) $ are the operators of annihilation and creation of a fermion, whereas $b_s (\bb p)$ and $b^\dag_s (\bb p) $ are the operators of annihilation and creation of an antifermion. 

The basis bispinors $u_s (\varepsilon (\bb p), \bb p),  \bar u_s (\varepsilon (\bb p), \bb p), v_s (\varepsilon (\bb p), \bb p),  \bar  v_s (\varepsilon (\bb p), \bb p)$ are (see Appendix A)

\begin{eqnarray}
 \label{eq6}
u_s(\varepsilon, \bb p) =
 \begin{pmatrix}
\sqrt{(p\sigma_-)}\xi_s\\
-i  \sqrt{(p\sigma_+)}\xi_s\\
\end{pmatrix}, ~~~v_s (\varepsilon, \bb p) =
 \begin{pmatrix}
\sqrt{(p\sigma_+)}\eta_s\\
i  \sqrt{(p\sigma_-)}\eta_s\\
\end{pmatrix}
\end{eqnarray}

\begin{eqnarray}
 \label{eq7}
\bar u_s(\varepsilon, \bb p) =( \xi_s^\dag i\sqrt{(p\sigma_+)} ;  \xi_s^\dag \sqrt{(p\sigma_-)}), ~~\bar v_s(\varepsilon, \bb p) =( -\eta_s^\dag i\sqrt{(p\sigma_-)} ;  \eta_s^\dag \sqrt{(p\sigma_+)}),
\end{eqnarray}
where $\xi_s$ and $\eta_s$ are two-component spinors, normalized by the condition $\xi^\dag_s\xi_s=\eta^\dag_s\eta_s=1$, $\sigma_\pm = (1, \pm \bb \sigma)$.

\subsubsection{Bispinor spaces}
 
 The direct derivations show that  the key Dirac scalar products of the basis spinors $u_s(\varepsilon, \bb p)$  and  $v_s(\varepsilon, \bb p)$   are equal to zero

\begin{subequations}\label{eq8}
\begin{eqnarray}
 \label{eq8a}
&&\bar u_s(\varepsilon, \pm\bb p) u_s(\varepsilon, \pm \bb p) =0 , \\\label{eq8b}
&& \bar v_s(\varepsilon,\pm\bb p)  v_s(\varepsilon, \pm\bb p)=0 ,\\\label{eq8c}
&&\bar u_s(\varepsilon,  \mp \bb p) v_s(\varepsilon, \pm \bb p) = \bar v_s(\varepsilon, \mp\bb p)  u_s(\varepsilon, \pm\bb p)=0.
\end{eqnarray}
\end{subequations}

The scalar products given by Eqs.(\ref{eq8a}),(\ref{eq8b}) show that the bispinors and co-bispinors (the Dirac conjunctive bispinors) are always orthogonal that becomes difficult in them normalization as compared with the case of the  standard Dirac fermions . A solution of this problem can be realized by extending the Dirac bispinor space up to the Clifford space\cite{Hod}, introducing the Clifford geometry product in the space, which is the unification of the Dirac bispinor $u$  and Dirac co-bispinor $\bar u$ spaces, since these  are no intersecting due to  Eqs.(\ref{eq8a}),(\ref{eq8b}).
In the case of the $u$-bispinors this extension means

  \begin{eqnarray}
 \label{eqa9a}
\bar u  \cdot  u= \bar u   u + \bar u  \wedge  u, 
\end{eqnarray}
where the first term is the standard scalar product, whereas the second one is  the external product.  We define this external product so that

   \begin{subequations}
  \begin{eqnarray}
 \label{eqa9b}
 && \bar u  \wedge  u =  i ( \bar u \gamma^5    u),\\
 && u  \wedge  \bar u  = i ( \bar u \gamma^5    u)^T, 
\end{eqnarray}
 \end{subequations}
where the same argument of $u$ and $\bar u$ is assumed.
The direct derivations show that such a definition leads to  anti-commutatibelity for  the external product 
  \begin{subequations}
  \begin{eqnarray}
 \label{eqa9b}
 &&u  \wedge  \bar u  +  u  \wedge  \bar u=0,\\
&& v \wedge  \bar v  +  v  \wedge  \bar v=0. 
\end{eqnarray}
 \end{subequations}

When $u$ and $v$ are given by Eqs.(\ref{eq6}),(\ref{eq7}) we have

 \begin{subequations}
\begin{eqnarray}
 &&\bar u (\bb p)  \wedge u( \bb p)  =2m,\\
&&\bar  v(\bb p) \wedge v (\bb p) =-2m, \\
&&\bar u (\bb p) \wedge v (-\bb p)  =0,\\
&&\bar v (\bb p) \wedge u (-\bb p)  =0. 
\end{eqnarray}
 \end{subequations}
Thus, the normalization of the basis Dirac bispinors can be carried out by means of the area  of the space, which is confined by the bispinors $u( v)$ and $\bar u(\bar v)$ in the Clifford geometrical space introduced above. This  area turns out to be  equal to $2m(-2m)$.

\subsubsection{Spin sums}

The explicit form of the basis bispinors  given by Eqs. (\ref{eq6}), (\ref{eq7}) allows us to obtain the spin sums which play a key role in the Feynman diagram calculations.   The direct derivations give

\begin{subequations}
\begin{eqnarray}
 &&\sum \limits_s^{} u_s(\varepsilon, \bb p) \bar u_s(\varepsilon, \bb p) = \gamma p - im \gamma^5, \label{eq7aa-1}\\
&&\sum \limits_s^{} v_s(\varepsilon, \bb p) \bar v_s(\varepsilon, \bb p) = \gamma p + im \gamma^5, \label{eq7aa-2}
\end{eqnarray}
\end{subequations}
that strongly differs from the standard spin sums\cite{Pes95}.

\subsection{Quantized anomalous fermion fields}

  Eq.(\ref{eq3a}) allows us to write down the Hamiltonian of  fermoins which is (see \cite{Pes95})
  
  \begin{eqnarray}
 \label{eq9}
&&H=\int {d^3\bb x} ~\psi^\dag (x) \left( - i \bb \alpha  \nabla + i m \gamma^0\gamma^5 \right)  \psi (x), 
\end{eqnarray}
where $x = (x^0, \bb x)$.

After the direct derivation, using the solutions given be Eqs.(\ref{eq4}),(\ref{eq5}), we obtain that the vacuum expectation of $H$ is

\begin{eqnarray}
 \label{eq10}
&&E = <0|H|0> =\sum_{s,s'} \int  \frac{d^3\bb p}{(2\pi)^3}\frac{d^3\bb p'}{(2\pi)^3}(<0| ( a^\dag_{s'} (\bb p') a_s (\bb p)|0> - <0|b_{s'} (\bb p') b^\dag_s (\bb p)|0>  )  \varepsilon (\bb p)
\end{eqnarray}

Thus, in order to the energy $E$ be positive definite value, the anticommutation relations have to be fulfilled

\begin{subequations}
 \begin{eqnarray}
 \label{eq11}
&& a_s (\bb p)  a^\dag_{s'} (\bb p') + a^\dag_{s'} (\bb p')a_{s} (\bb p) = (2\pi)^3 \delta(\bb p -\bb p') \delta_{ss'}\\
&&  b_s (\bb p)  b^\dag_{s'} (\bb p') + b^\dag_{s'} (\bb p')b_{s} (\bb p) = (2\pi)^3 \delta(\bb p -\bb p') \delta_{ss'}.
\end{eqnarray}
\end{subequations}
Then, as a result, we get

\begin{eqnarray}
 \label{eq12}
&&E= \sum_{s} \int  \frac{d^3\bb p}{(2\pi)^3}(n_s (\bb p)+{\bar n}_s (\bb p)-1 )  \varepsilon (\bb p),
\end{eqnarray}
that corresponds to the standard quantization of fermion fields. In Eq.(\ref{eq12})  the occupancy numbers of fermions  $n_s (\bb p)=<0| ( a^\dag_{s} (\bb p) a_s (\bb p)|0> $ and antifermions${\bar n}_s (\bb p)=<0| ( b^\dag_{s} (\bb p) b_s (\bb p)|0> $ are introduced.
The analogical calculations leads to the formulas for a momentum and a charge of the fermion-antifermion system which coincide with Eqs.(3.108),(3.113)\cite{Pes95}.

\subsection{Lagrangian of anomalous fermions }

Eq.(\ref{eq3a}),(\ref{eq3b}) allow us to derive the action integral ${\cal A}$ and the Lagrangian ${\cal L}$ which  govern the dynamics of the anomalous fermions. They   obviously have a form

\begin{eqnarray}
 \label{eq13}
&&{\cal A} = \int d^4x {\cal L} =\int d^4 x \left\{ \frac{i}{2} (\bar\psi (x) ( \gamma^\mu \partial_\mu - \gamma^5m) \psi (x))-\frac{i}{2}  (\bar\psi (x) (\gamma^\mu {\overleftarrow \partial}_\mu + \gamma^5m) \psi (x))\right\}.
\end{eqnarray}

This Lagrangian generates the standard $J^\mu = \bar\psi (x) \gamma^\mu \psi (x) $ and axial currents $J_a^\mu = \bar\psi (x) \gamma^\mu \gamma^5 \psi (x) $ which satisfy the conservation laws

 \begin{subequations}
\begin{eqnarray}
 &&\partial_\mu J^\mu = 0, \label{eq14a}\\
&&\partial_\mu J_a^\mu =-2m \bar\psi (x)  \psi (x). \label{eq14b}
\end{eqnarray}
\end{subequations}
We note that the non-conservation of the axial current, which  is given by Eq.(\ref{eq14b}), strongly differs from the case of the standard Dirac fermions\cite{Itz,Pes95}. Therewith, the axial current is conserved in the chiral limit, as it takes place in the case of free Dirac fermions.

\subsection {Discrete symmetries}

We consider the parity ($P$), time traversal ($T$) and charge conjugation ($C$) of the anomalous fermion field. In terms of the solutions given by Eqs.(\ref{eq4}),(\ref{eq5}) these symmetries  mean transformations of the operators of creation and annihilation of fermions as follows\cite{Pes95}

 \begin{subequations}
\begin{eqnarray}
 \label{eq15}
&&P: \longrightarrow ~~~ P a_s(\bb p) P =\zeta_a s_a(-\bb p), ~~~~P b_s(\bb p) P =\zeta_b b_s (-\bb p);  \\
&&T: \longrightarrow ~~~ T a_s(\bb p) T =a_{-s}(-\bb p), ~~~~T b_s(\bb p) T = b_{-s} (-\bb p);\\
&& C: \longrightarrow ~~~ C a_s(\bb p) P =b_a(\bb p), ~~~~P b_s(\bb p) P =a_s (\bb p).\label{eq15b}
\end{eqnarray}
\end{subequations}

Taking into account that $u_s (\bb p) $ and $ v_s (\bb p) $ are given by Eqs.(\ref{eq6}),(\ref{eq7}) rather that by Eqs.(3.50),(3.62)\cite{Pes95},  the equations transforming $\psi (x) $ and $\bar \psi (x)$ are modified as compared with Ref.\cite{Pes95}, and take a form

 \begin{subequations}
\begin{eqnarray}
 \label{eq16}
&& P\psi(x)P =i \gamma^0 \gamma^5 \psi(-x)\\
&& P\bar \psi(x)P =   -i \bar \psi (-x)\gamma^0   \gamma^5 
\end{eqnarray}
\end{subequations}

 \begin{subequations}
\begin{eqnarray}
 \label{eq17}
&&  T\psi(t, \bb x )T =i\gamma^1 \gamma^3\gamma^5  \psi(-t, \bb x )\\
&& T\bar \psi(t, \bb x )T = +i  \bar \psi(-t, \bb x )\gamma^5 \gamma^3 \gamma^1
\end{eqnarray}
\end{subequations}

 \begin{subequations}
\begin{eqnarray}
 \label{eq18}
&& C \psi(x )C=( i\bar \psi (x) \gamma^0 \gamma^2 \gamma^5 )^T\\
&& C \bar \psi(x )C =( i\gamma^0  \gamma^2 \gamma^5  \psi (t x ))^T \label{eq18b}
\end{eqnarray}
\end{subequations}

Then, in terms  of Eqs.(\ref{eq16})-(\ref{eq18b}) we derive Table I which demonstrates changing the bi-linear combinations of the the fields $\bar \psi (x), \psi (x)$ given by  Eqs.(\ref{eq4}),(\ref{eq5}) under the $P,T, C$ transformations.

\begin{table}[H]
\caption{ Bi-linear combination under P,T,C transformations. \\ $(-1)^\mu \equiv 1 $ at $  \mu=0$, and $(-1)^\mu \equiv-1$ at $  \mu=1,2,3$.\\The signs in square brackets correspond to the standard fermion case\cite{Pes95}.}
\begin{center}
\begin{tabular}{|c|c|c|c|c|}
\hline  & & & &\\
~~&  $\bar \psi \psi $ & $i\bar \psi \gamma^5 \psi$ & $ \bar \psi \gamma^\mu \psi $ & $  \bar \psi \gamma^\mu  \gamma^5   \psi $ \\
  \hline
P &  $-[+]$&$+ [-]$& $(-1)^\mu[(-1)^\mu]$ & $(-1)(-1)^{\mu}[(-1)(-1)^{\mu}]$\\ 
  \hline
T& $-[+]$& $+ [-]$& $(-1)^\mu[(-1)^\mu]$ & $(-1)^{\mu}[(-1)^{\mu}]$\\  
  \hline
 C  & $+[+]$&$ + [+]$& $+[-]$ & $-[+]$ \\
   \hline
CPT   &$+[+]$& $+ [+]$& $+[-]$ & $+[-]$ \\
   \hline
 \end{tabular}
\end{center}
\end{table}

It is in  particular seen from Tab.1,  that the scalar and pseudo scalar combinations  of the anomalous fermion fields are transformed by P and T transformations  with the opposite sign as compared with case of standard Dirac fermions\cite{Pes95}.

\section{Conclusion}

The anomalous   fermions with spin$1/2$ which are governed by  the  new Dirac-like equation are  studied. The considered fermion fields are strongly different from  the classic Dirac fermions that  manifests itself both in properties of basis bispinors and in the C,P,T symmetries. These specifics of the obtained fields allows us to consider them as candidates to the particles consisting of   dark matter.   It, particularly, follows from the spin sums given by Eqs.(\ref{eq7aa-1}),(\ref{eq7aa-2}),  which contain the terms being  proportional to $\gamma^5$. The presence of  such a term  likely  has to change the probabilities of weak processes, such as the muon decay, for example,  compared  with the case of the classic Dirac fermions.

\appendix{}

\section{Solution of the motion equation for anomalous fermions }

We go to the momentum representation in  Eq.(\ref{eq3a}), and write it  in the explicit form, using Eq.(\ref{eq2}). As a result, we get

\begin{eqnarray}
 \label{a1}
 \begin{pmatrix}
im  &\omega+\bb p \bb \sigma\\
\omega - \bb p \bb \sigma  & -im   \\
\end{pmatrix} 
 \begin{pmatrix}
\varphi (p) \\
\chi (p)  \\
\end{pmatrix} =0,
\end{eqnarray}
where $p=(\omega, \bb p)$ determine a point in the momentum space, $\bb \sigma$ are the Pauli matrices,$\varphi (p)$ and $\chi (p)$ are spinors which are the components of a bispinor $\psi (p)$. To take place a non-trivial solutions of  Eq.(\ref{a1}) the following is demanded

\begin{eqnarray}
 \label{a2}
&& m^2 - (\omega+\bb p \bb \sigma)(\omega-\bb p \bb \sigma) =0, 
\end{eqnarray}
that leads to 

\begin{eqnarray}
 \label{a3}
&& \omega = \pm \varepsilon = \pm \sqrt{ m^2 +\bb p^2}.
\end{eqnarray}
Two signs in the latter equation dictates to consider two cases. We note that the four-vector $p=(\varepsilon (\bb p)) $ is on-shell in the both cases.

i) Let us consider $\omega = \varepsilon (\bb p)=  \sqrt{ m^2 +\bb p^2}$. We introduce the following notations

\begin{eqnarray}
 \label{a4}
&& \sigma_+ =(1, \bb \sigma),~~~~~ \sigma_- =(1, -\bb \sigma),
\end{eqnarray}

Then, the solution of Eq. (\ref{a1}) is 
\begin{eqnarray}
 \label{a5}
u_s(\varepsilon, \bb p) =
 \begin{pmatrix}
\varphi \\
\chi   \\
\end{pmatrix} = \begin{pmatrix}
\sqrt{(p\sigma_-)}\xi_s\\
-i  \sqrt{(p\sigma_+)}\xi_s\\
\end{pmatrix},
\end{eqnarray}
where $\xi_s$ is the standard spinor\cite{Pes95}, normalized by the condition $\xi^\dag \xi=1$.
We directly obtain from Eq. (\ref{a5}) that

\begin{subequations}
\begin{eqnarray}
 \label{a6}
&& \bar u_s(\varepsilon, \bb p) =( \xi_s^\dag i\sqrt{(p\sigma_+)} ;  \xi_s^\dag \sqrt{(p\sigma_-)}),\\
&&\bar u_s(\varepsilon, \bb p) u_s(\varepsilon, \bb p)=0, \\
&& u^\dag_s(\varepsilon, \bb p) u_s(\varepsilon, \bb p)=2 \varepsilon (\bb p)
\end{eqnarray}
\end{subequations}

ii) Let $\omega$ be  $\omega = -\varepsilon (\bb p)= - \sqrt{ m^2 +\bb p^2}$. Then, Eq.(\ref{a1}) takes a form

\begin{eqnarray}
 \label{a7}
&& im \varphi  + (-\varepsilon+\bb p \bb \sigma)\chi =0\nonumber \\
&& (-\varepsilon-\bb p \bb \sigma)\varphi - im \chi =0.
\end{eqnarray}
The solution of Eq.(\ref{a7}) can be written as follows

\begin{eqnarray}
 \label{a8}
v (\varepsilon, \bb p) =
 \begin{pmatrix}
\varphi \\
\chi   \\
\end{pmatrix} = \begin{pmatrix}
\sqrt{(p\sigma_+)}\eta_s\\
i  \sqrt{(p\sigma_-)}\eta_s\\
\end{pmatrix}, 
\end{eqnarray}
where where $\eta_s$ is the standard spinors\cite{Pes95}, normalized by the condition $\eta^\dag \eta=1$. We derive from Eq.(\ref{a8}) that

\begin{subequations}
\begin{eqnarray}
 \label{a9}
&&\bar v_s(\varepsilon, \bb p) =( -\eta_s^\dag i\sqrt{(p\sigma_-)} ;  \eta_s^\dag \sqrt{(p\sigma_+)}),\\
&&\bar v_s(\varepsilon, \bb p) v_s(\varepsilon, \bb p)=0, \\
&& v^\dag_s(\varepsilon, \bb p) v_s(\varepsilon, \bb p)=2 \varepsilon (\bb p)
\end{eqnarray}
\end{subequations}

Using the basis bispinors $ u_s(\varepsilon, \bb p)$  and $ v_s(\varepsilon, \bb p) $, which are given by Eqs.(\ref{a5}),(\ref{a8}),  the general solution of Eqs.(\ref{eq3a}),(\ref{eq3b}) can be written as Eqs.(\ref{eq4}),(\ref{eq5})\cite{Pes95}.

\end{document}